\documentclass[prc,twocolumn,showpacs,floatfix,superscriptaddress,nofootinbib]{revtex4}
\usepackage{graphicx}
\usepackage{amssymb}
\usepackage{dcolumn}
\usepackage{amsmath}
\usepackage{bm}
\usepackage{textcomp}
\usepackage{epstopdf}
\usepackage{color}
\usepackage{ulem}
\usepackage[toc,page,title,titletoc,header]{appendix}
\usepackage[colorlinks,
            linkcolor=blue,
            anchorcolor=blue,
            citecolor=blue
            ]{hyperref}
\usepackage{txfonts}
\usepackage{caption}

\begin{document}

\title{Universal scaling of conserved charge in the stochastic diffusion dynamics}

\begin{abstract}
In this paper, we explore the Kibble-Zurek scaling of the conserved charge, using the stachastic diffusion dynamics. After determining the characteristic scales $\tau_{\mbox{\tiny KZ}}$ and $l_{\mbox{\tiny KZ}}$ and properly rescaling the traditional correlation function and cumulant, we construct universal functions for both the two-point correlation function $C(y_1-y_2;\tau)$ and second-order cumulant $K(\Delta y,\tau)$ of the conserved charge in the critical regime, which are insensitive to the initial temperature and a parameter in the mapping between 3D Ising model and the hot QCD system near the critical point.
\end{abstract}

\author{Shanjin Wu}
\email{shanjinwu2014@pku.edu.cn}
\affiliation{Department of Physics and State Key Laboratory of Nuclear Physics and
Technology, Peking University, Beijing 100871, China}

\author{Huichao Song}
\email{huichaosong@pku.edu.cn}
\affiliation{Department of Physics and State Key Laboratory of Nuclear Physics and
Technology, Peking University, Beijing 100871, China}
\affiliation{Collaborative Innovation Center of Quantum Matter, Beijing 100871, China}
\affiliation{Center for High Energy Physics, Peking University, Beijing 100871, China}
\maketitle


\section{Introduction}\label{Intro}
One of the main goals of the Beam Energy Scan (BES) program~\cite{Aggarwal:2010cw,Mohanty:2011nm,Kumar:2013cqa,Adamczyk:2017iwn,Odyniec:2015iaa} at Relativistic Heavy Ion Collider (RHIC) is to probe the phase structure of Quantum Chromodynamics (QCD) matter and to search the critical point~\cite{Stephanov:1999zu,Stephanov:1998dy,Stephanov:2004wx,Stephanov:2007fk,Asakawa:2015ybt,Luo:2017faz,Li:2017ple,Chen:2018vty,Li:2018ygx,Fu:2015amv}, the endpoint of the 1-st order phase transition boundary of the QCD phase diagram~\cite{Stephanov:1998dy,Stephanov:2004wx,Stephanov:2007fk,Klevansky:1992qe,Fukushima:2003fw,Fu:2007xc,Jiang:2013yoa,Roberts:1994dr,Qin:2010nq,Berges:2000ew}. At the critical point, the thermal medium is strongly correlated with diverge fluctuations of various variables~\cite{Stephanov:1999zu,Stephanov:1998dy,Stephanov:2004wx, Stephanov:2007fk}. It was also found that the skewness $S$ and kurtosis $\kappa$ of the net protons diverge with the correlation length by $\xi^{4.5}$ and $\xi^{7}$, respectively~\cite{Stephanov:2008qz}. In BES experiment, event-by-event multiplicity fluctuations of net protons and net charges have been systematically measured at different collision energies~\cite{Aggarwal:2010wy,Adamczyk:2013dal,Adamczyk:2014fia,Thader:2016gpa,Luo:2015ewa}. It was found that the kurtosis $\kappa$ of net protons, presents a non-monotonic behavior and largely deviates from the Poisson baseline at lower collision energies, indicating the potential of discovery the critical point~\cite{Stephanov:2011pb,Luo:2015ewa}.

Recently, it was realized that the non-equilibrium effects are significant for an expanding medium near the critical point~\cite{Stephanov:2009ra,Son:2004iv,Ling:2015yau,Stephanov:2017ghc,Mukherjee:2015swa, Mukherjee:2016kyu,Brewer:2018abr,Akamatsu:2018vjr,Rajagopal:2000,Paech:2003fe,Nahrgang:2011mg,Nahrgang:2018afz, Nonaka:2004pg,Asakawa:2009aj,Sakaida:2017rtj,Jiang:2015hri,Jiang:2017mji}. In particular, the critical slowing down effects largely influence the non-equilibrium fluctuations, which even reverse the signs of skewness $S$ and kurtosis $\kappa$ compared with the equilibrium values~\cite{Mukherjee:2015swa,Jiang:2017mji}. It was also argued that the soft mode of the critical point is a diffusive mode, which is a combination of the order parameter field and the conserved quantities~\cite{Son:2004iv}. Recently, diffusion dynamics near the critical point  have been developed~\cite{Sakaida:2017rtj,Nahrgang:2018afz}, which showed that the second order cumulant of the conserved charge presents non-monotonic behavior with the change of the rapidity window~\cite{Sakaida:2017rtj}.

For the dynamical model calculations near the critical point, the non-equilibrium fluctuations are non-universal, which depend on various free parameters, such as the relaxation time and the mapping from the 3D Ising model to the hot QCD medium, etc. On the other hand, within the framework of Kibble-Zurek Mechanism (KZM), one can construct some universal variables near the critical point that are insensitive to some non-universal factors~\cite{Mukherjee:2016kyu,Chandran:2012,Kolodrubetz:2012,Francuz:2015zva,Nikoghosyan:2013fqa,Wu:2018twy,Akamatsu:2018vjr}. The key point of the KZM is that, due to the critical slowing down effects, the systems inevitably get out-of-equilibrium near the critical point, after which these ``frozen'' systems have correlated regions with characteristic scales, leading to various universal variables. The KZM was first introduced by Kibble in cosmology~\cite{Kibble:1976} and then extended by Zurek to the condensed matter physics~\cite{Zurek:1985}. In relativistic heavy ion collisions, the KZM was first studied in Ref.~\cite{Mukherjee:2016kyu}, which constructed universal functions of the order parameter field that are insensitive to the relaxation time and the evolving trajectory of the system. In Ref.~\cite{Wu:2018twy}, we have investigated the Kibble-Zurek scaling for both the order parameter field and the multiplicity fluctuations of net protons, using the Langevin dynamics of model A.  We found that, compared with the original fluctuations of net protons, the oscillating behavior of the constructed approximately universal functions are strongly suppressed.

In this paper, we investigate the critical universal scaling of the conserved charge within the framework of model B. As mentioned above, the soft mode near the QCD critical point is a diffusive mode, which is a linear combination of the order parameter field and the conserved quantities. {Moreover, the conserved quantities directly related to the possible experimental observable}.  Comparing with our early work~\cite{Wu:2018twy} which only considers the non-conserved order parameter field, this paper explores the possible universal scaling for fluctuation of conserved charge using the stochastic diffusion equation (SDE). We will demonstrate that the constructed universal functions for the two-point correlation function and the second-order cumulant of conserved charge are insensitive to the non-universal factors of two cases 1) the evolving hot medium with different strength of critical component $c_c$,  a parameter in the mapping from 3D Ising to QCD critical point; 2) the evolving system with different initial temperature $T_0$. Note that, in this paper, we restrict our attention to the possibility of constructing the universal functions for the diffusion dynamics near the critical point, which we only focus on the 1+1-dimensional system with the Bjorken approximation. For the realistic universal observables that might be associated with experimental measurements, one needs to at least numerically simulate the 3+1 dimensional diffusion dynamics and consider the higher-order cumulants of fluctuations. This requires high statistical runs and a large amount of computing resources, which we would like to leave it to future study.

The paper is organized as follows: Sec.~\ref{Model} briefly reviews the dynamics of conserved charge near the critical point based on the stochastic diffusion equation. In Sec.~\ref{Univ}, we construct the universal functions for two-point correlation function and the second-order cumulant. Sec.~\ref{Results} presents and discusses the main results of the constructed universal functions. Sec.~\ref{summary} summarizes and concludes this paper.

\section{Dynamics of conserved charge}\label{Model}

\subsection{Stochastic diffusion equation}

{For a dynamical model near the critical point, the slow modes are the relevant and essential modes, which largely influence the critical behavior of the evolving system.} According to the classification of Ref.~\cite{Rev1977}, the critical dynamics of non-conserved and conserved order parameter field belong to model A and B, respectively. While, model H describes a system with a conserved order parameter field, conserved transverse momentum density, and nonzero Poisson bracket between the two. In general, it is believed that the dynamical system near QCD critical point lies in model H~\cite{Son:2004iv,Fujii:2003bz,Fujii:2004jt,Fujii:2004za}.  However, the related analysis or numerical implementation of model H is complicated, which have not been fully developed. For simplicity, our previous work~\cite{Wu:2018twy} only focused on the dynamics and universal scaling of the non-conserved order parameter field within the framework of model A. Recently, Ref.~\cite{Sakaida:2017rtj} has developed the stochastic diffusion dynamics of the conserved charge for model B, {which demonstrated that the two-point correlation function and cumulant behave non-monotonically with the change of the rapidity interval and window, respectively}. In this paper, we will explore the universal behavior of the conserved charge based on the stochastic diffusion equation described in Ref.~\cite{Sakaida:2017rtj}.

For simplicity, we focus on 1+1-dimensional  evolution of the conserved charge density $n(y,\tau)$ with the proper time $\tau=\sqrt{t^2-z^2}$ and the spacetime rapidity $y=\tanh^{-1}(z/t)$ for a boost-invariant Bjorken system.  The related stochastic diffusion equation is~\cite{Sakaida:2017rtj}:
\begin{align}\label{SDE}
\frac{\partial}{\partial \tau} \delta n(y,\tau) = D_y(\tau) \frac{\partial^2}{\partial^2 y} \delta n(y,\tau) + \frac{\partial }{\partial y} \zeta(y,\tau)
\end{align}
Here $\delta n(y,\tau)=n(y,\tau)-\langle n(y,\tau)\rangle$, and $\langle \cdots\rangle$ denotes the event average.  The diffusion coefficient $D_y(\tau)$ is related to the Cartesian one $D_C(\tau)$ with $D_y(\tau)=D_C(\tau)\tau^{-2}$. The noise $\zeta(y,\tau)$ satisfies the fluctuation-dissipation theorem:
\begin{align}
\begin{aligned}
&\langle \zeta(y,\tau) \rangle =0,\\
&\langle \zeta(y_1,\tau_1)\zeta(y_2,\tau_2)\rangle = 2\chi_y(\tau) D_y(\tau) \delta (y_1-y_2) \delta (\tau_1-\tau_2),
\end{aligned}
\end{align}
where $\chi_y(\tau)$ is the susceptibility of the conserved charge per unit rapidity, related to the Cartesian one $\chi_C(\tau)$ with $\chi_y(\tau)/\tau=\chi_C(\tau)$. For notational convenience, the subscripts of the diffusion coefficient and susceptibility in the following part of this paper are dropped, which are denoted as $D(\tau)=D_y(\tau)$ and $\chi(\tau)=\chi_y(\tau)$, respectively.

After solving the SDE~(\ref{SDE}), one could obtain the correlation function
\begin{align}\label{Correl}
\begin{aligned}
C(y_1,y_2;\tau)&\equiv \langle \delta n(y_1,\tau)\delta n(y_2,\tau) \rangle\\
&= \chi(\tau) \delta(y_1-y_2)\\
 &\qquad- \int^\tau_{\tau_0} d\tau' \chi'(\tau') G(y_1-y_2;2d(\tau',\tau)),
\end{aligned}
\end{align}
where $\chi'(\tau)=d\chi(\tau)/d\tau$.
Here the normalized Gauss distribution is:
\begin{align}
G(\bar{y};d) \equiv \frac{1}{\sqrt{\pi}d} e^{-\bar{y}^2/d^2},
\end{align}
and
\begin{align}
d(\tau_1,\tau_2)\equiv\left[2\int^{\tau_2}_{\tau_1}d\tau' D(\tau')\right]^{1/2}
\end{align}
{represents the diffusion ``length'' in rapidity space} from $\tau_1$ to $\tau_2$ with $\tau_1\leq \tau_2$.

The amount of the charge deposed within a finite rapidity window $\Delta y$ at mid-rapidity and at a proper time $\tau$ can be calculated as:
\begin{align}
Q_{\Delta y}(\tau) \equiv \int^{\Delta y/2}_{-\Delta y/2} dy n(y,\tau).
\end{align}
Correspondingly, the second-order cumulant of $Q_{\Delta y}(\tau)$ takes the following form:
\begin{align}\label{Cumul}
\begin{aligned}
K(\Delta y,\tau)&\equiv  \langle \delta Q_{\Delta y}(\tau)^2\rangle/\Delta y\\
&=\frac{1}{\Delta y}\int^{\Delta y/2}_{-\Delta y /2} dy_1dy_2 \langle \delta n(y_1,\tau) \delta n(y_2,\tau)\rangle\\
&=\chi(\tau)- \int^\tau_{\tau_0} d\tau' \chi'(\tau') F\left(\frac{\Delta y}{2d(\tau',\tau)}\right),
\end{aligned}
\end{align}
where
\begin{align}
F(X)\equiv \frac{2}{\sqrt{\pi}} \int^X_0 dz\left(1-\frac{z}{X}\right)e^{-z^2}.
\end{align}
For the detailed derivation, please refer to Appendix.~\ref{Appendix}.

Note that the SDE~(\ref{SDE}) used in the present study only considers the two-point interaction and neglect the higher order contributions. The advantage of such simplification is that it can be analytically solved, as shown in Eqs. (\ref{Correl}) and (\ref{Cumul}). It is adequate for our first attempt to study the Kibble-Zurek scaling for the two-point correlations in the diffusion dynamics without further considering the higher order cumulants.

\subsection{Parametrizing the susceptibility $\chi(\tau)$ and diffusion coefficient $D(\tau)$}

Both the correlation function (\ref{Correl}) and the cumulant (\ref{Cumul})  depend on the susceptibility $\chi(\tau)$ and diffusion coefficient $D(\tau)$, which needs some additional parametrizations.
In general, the susceptibility $\chi$ and diffusion coefficient $D$ include both the singular parts $\chi^{cr}$, $D^{cr}_C$ and the regular parts $\chi^{reg}$, $D^{reg}_C$, respectively~\cite{Sakaida:2017rtj}. As the system evolves near the critical point,  the singular contributions become dominant.  We thus neglect the regular parts to simplify the following study of the Kibble-Zurek scaling. The susceptibility $\chi(\tau)$ and diffusion coefficient $D(\tau)$ with only the singular parts are then written as:
\begin{align}
&\chi(T)=\chi^{cr}(T),\\
&D(T)=D^{cr}_C/\tau^2.
\end{align}

Here, we construct the singular part $\chi^{cr}$ and $D^{cr}_C$ through a mapping between the hot QCD matter and the 3D Ising model. In the linear parametric model~\cite{Justin:2001,Schofield:1969}, the magnetization of the 3D Ising systems is parameterized with two variables $R$ and $\theta$:
\begin{align}\label{Magnet}
M(R,\theta) = m_0 R^{1/3}\theta,
\end{align}
where the reduced temperature $r$ and the dimensionless magnetic field $H$ are expressed as:
\begin{align}
&r(R,\theta)=R(1-\theta^2),\\
&H(R,\theta)=h_0 R^{5/3}\theta(3-2\theta^2),
\end{align}
Here, we have adopted the values of the Ising critical exponents~\cite{Guida:1996ep}, and the normalization constants $m_0$ and $h_0$ are fixed by the conditions $M(r=-1,H=0^+)=1$ and $M(r=0,H=1)=1$. From Eq.~(\ref{Magnet}),  one could calculate the susceptibility of the 3D Ising model:
\begin{align}
\chi_M(r,H) = \frac{\partial M(r,H)}{\partial H} \Bigg|_r= \frac{m_0}{h_0} \frac{1}{R^{4/3}(3+2\theta^2)},
\end{align}

In the case of the hot QCD systems, the susceptibility $\chi^{cr}$  for the conserved charge satisfies a similar critical behavior near the critical point:
\begin{align}
\frac{\chi^{cr}(r,H)}{\chi^H} = c_c \chi_M(r,H) =c_c \frac{m_0}{h_0} \frac{1}{R^{4/3}(3+2\theta^2)},
\end{align}
where the dimensionless factor $c_c$ is treated as a free parameter. $\chi^H$ is the susceptibility in the hadronic medium, which can be absorbed by the definitions $C'(y_1-y_2;\tau)\equiv C(y_1-y_2;\tau)/\chi^H$ and $K'(\Delta y,\tau)\equiv K(\Delta y,\tau)/\chi^H$. In the following calculations, we will omit the prime to simplify the notation.

Considering that the evolving hot QCD system belongs to model H in the classification of Ref.~\cite{Rev1977}, we scale
the  diffusion coefficient $D^{cr}_C$ with the correlation length $\xi$ as: $D^{cr}_C\sim \xi^{-2-\chi_\eta+\chi_\lambda}$ with the exponents $\chi_\eta\simeq 0.04$ and $\chi_\lambda\simeq 0.916$~\cite{Rev1977}. The correlation length $\xi$ is connected to the susceptibility $\chi^{cr}$ as:
\begin{align}
\xi=\xi_0\left(\frac{\chi^{cr}}{\chi^H}\right)^{1/(2-\chi_\eta)},
\end{align}
Here we set $\xi_0=0.1$fm. Correspondingly, the parameterized $D^{cr}_C$ is:
\begin{align}
D^{cr}_C(r,H)=d_c \left[\frac{\chi^{cr}(r,H)}{\chi^H}\right]^{(-2+\chi_\eta+\chi_\lambda)/(2-\chi_\eta)}
\end{align}
where the constant $d_c=1$ fm, as used in Ref.~\cite{Sakaida:2017rtj}.

After the above parametrization, the susceptibility $\chi^{cr}(T,\mu)$ and diffusion coefficient $D^{cr}_C(T,\mu)$ as functions of temperature $T$ and chemical potential $\mu$ can be obtained from  $\chi^{cr}(r,H)$ and $D^{cr}_C(r,H)$ with the following linear mapping~\cite{Sakaida:2017rtj,Mukherjee:2015swa}
\begin{align}
\frac{T-T_c}{\Delta T} = \frac{H}{\Delta H},\qquad \frac{\mu-\mu_c}{\Delta \mu} = -\frac{r}{\Delta r},
\end{align}
where $r$ is treated as a free parameter to simulate the change of $\mu$~\cite{Sakaida:2017rtj}. The critical temperature is set to $T_c=160$ MeV and the width of the critical region is set to $\Delta T/\Delta H=10$ MeV.

Again, we only focus on an evolving system in 1+1-dimension with Bjorken expansion. We assume the heat bath is evolving along a trajectory with fixed $r$ and the temperature $T$ dropping down with the proper time $\tau$ as~\cite{Mukherjee:2015swa}:
\begin{align}\label{Bjorken}
T(\tau)=T_0\left(\frac{\tau_0}{\tau}\right)^{c^2_s}
\end{align}
where the speed of sound is taken as: $c^2_s=0.15$. The values of the initial time $\tau_0$ and the corresponding temperature $T_0$  will be explained in Sec.~\ref{Results}.

\section{{The Kibble-Zurek Scaling}}\label{Univ}

The correlation (\ref{Correl}) and cumulant (\ref{Cumul}) obtained from solving SDE~(\ref{SDE}) are non-universal and sensitive to some inputs in the parametrization of $\chi(\tau)$ and $D(\tau)$, such as the strengths of the critical component $c_c$, the initial temperature $T_0$, etc. In Refs.~\cite{Mukherjee:2016kyu} and ~\cite{Wu:2018twy}, the universal functions have been constructed within the framework of Kibble-Zurek Mechanism for model A,  {that involves with the evolving non-conserving order parameter field near the critical point}. In this section,  we will study the possible universal behavior of the correlation function (\ref{Correl}) and cumulant (\ref{Cumul}) for the evolving conserved charge of model B.

For a dynamical system near the critical point, there are two competitive time scales, the relaxation time $\tau_{\mbox{\tiny rel}}$ that describes the time for the system to equilibrate and the quench time $\tau_{\mbox{\tiny quench}}$ that characterizes the changing rate of the external potential.

Bjorken expansion of the hot medium Eq.~(\ref{Bjorken}) introduces the variation of the susceptibility $\chi(\tau)$ and diffusion coefficient $D(\tau)$ with which the quench time can be calculated as:
\begin{align}
\tau_{\mbox{\tiny quench}} = \left|\frac{\xi(\tau)}{\partial_\tau \xi(\tau)}\right|.
\end{align}

For a diffusion system near the critical point, the relaxation time of the two-point correlation function takes the form $ \tau_{\mbox{\tiny rel}} = [2 D(\tau) q^2]^{-1}$ for a  particular mode $q$. For the slow modes with $q\ll \xi^{-1}$, the relaxation time is large  { comparing to $\tau_{\mbox{\tiny quench}}$}, which leads to these modes out-of-equilibrium as the system evolves near the critical point. For the fast modes with $q\gg \xi^{-1}$, the relaxation times are small, which corresponds to fast enough equilibration even near the critical point.
In this work, we focus on the mode with $q\xi=1$ and the relaxation time is given by:
\begin{align}\label{rel}
\tau_{\mbox{\tiny rel}} = \frac{\xi^2}{2 D(\tau)}.
\end{align}

\begin{figure}[tbp]
\center
\includegraphics[width=3.0 in]{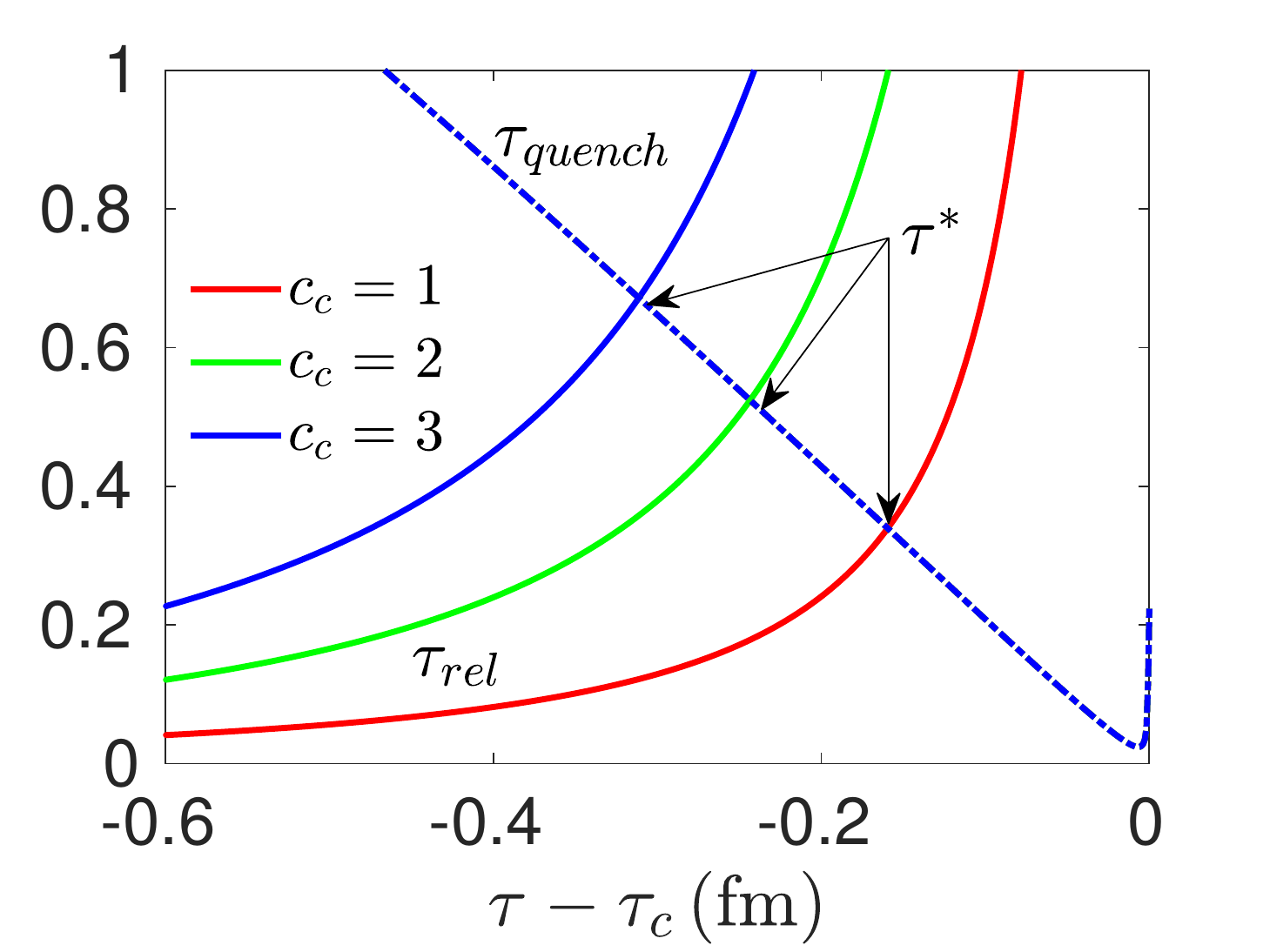}
\caption{(Color online)  Time evolution of the quench time $\tau_{\mbox{\tiny quench}}$ and the
 relaxation time $\tau_{\mbox{\tiny rel}}$ with different $c_c$. The location of the proper time $\tau^*$ is computed from $\tau_{\mbox{\tiny rel}}(\tau^*)=\tau_{\mbox{\tiny quench}}(\tau^*)$.}
\label{KZcc}
\end{figure}

Note that the relaxation time $\tau_{\mbox{\tiny rel}}$ strongly enhances as the system {cools down} to the critical point and the quench time $\tau_{\mbox{\tiny quench}}$ continuously decreases. As a consequence, there exists a point $\tau^*$, where the relaxation time equals to quench time, after which the system becomes out-of equilibrium with the formation of correlated patches. According to the Kibble-Zurek Mechanism, the characteristic time scale $\tau_{\mbox{\tiny KZ}}$ and length scale $l_{\mbox{\tiny KZ}}$ are determined by $\tau^*$ with~\cite{Mukherjee:2016kyu}:
\begin{align}\label{KZ}
  \tau_{\mbox{\tiny KZ}} = \tau_{\mbox{\tiny eff}}(\tau^*) = \tau_{\mbox{\tiny quench}} (\tau^*),
\qquad  l_{\mbox{\tiny KZ}} = \xi_{\mbox{\tiny eq}}(\tau^*).
\end{align}
In Fig.~\ref{KZcc}, we plot the relaxation time $\tau_{\mbox{\tiny rel}}$ and quench time $\tau_{\mbox{\tiny quench}}$ as functions of $\tau-\tau_c$, where $\tau_c$ is the time when the temperature of the system hits the critical temperature $T_c$. It shows that the relaxation time $\tau_{\mbox{\tiny rel}}$ increases and the quench time $\tau_{\mbox{\tiny quench}}$ decreases as the system approaches to the critical point, and the proper time $\tau^*$ can be determined by Eq.~(\ref{KZ}).

\begin{figure*}[tbp]
\center
\includegraphics[width=3.3 in]{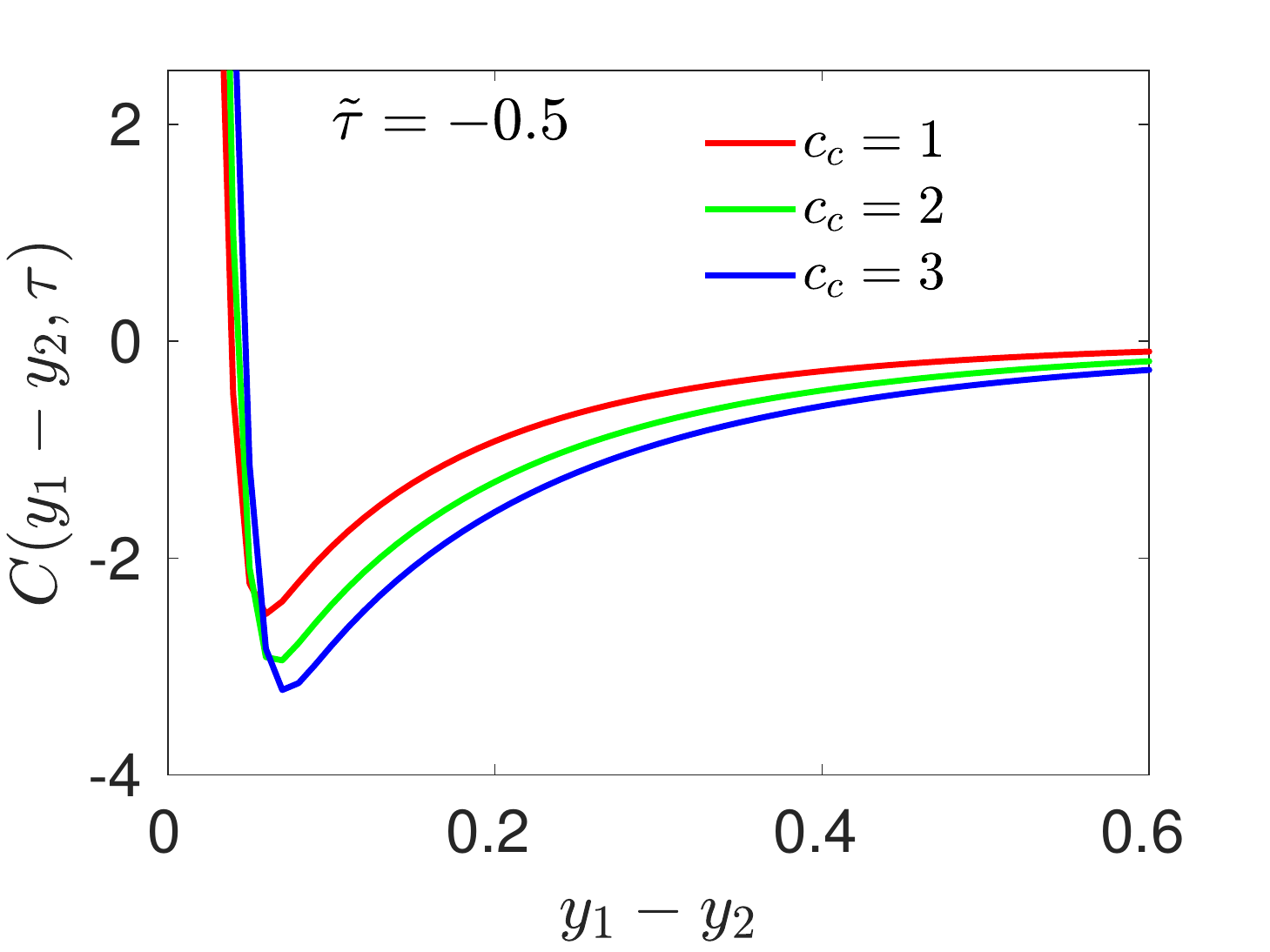}
\includegraphics[width=3.3 in]{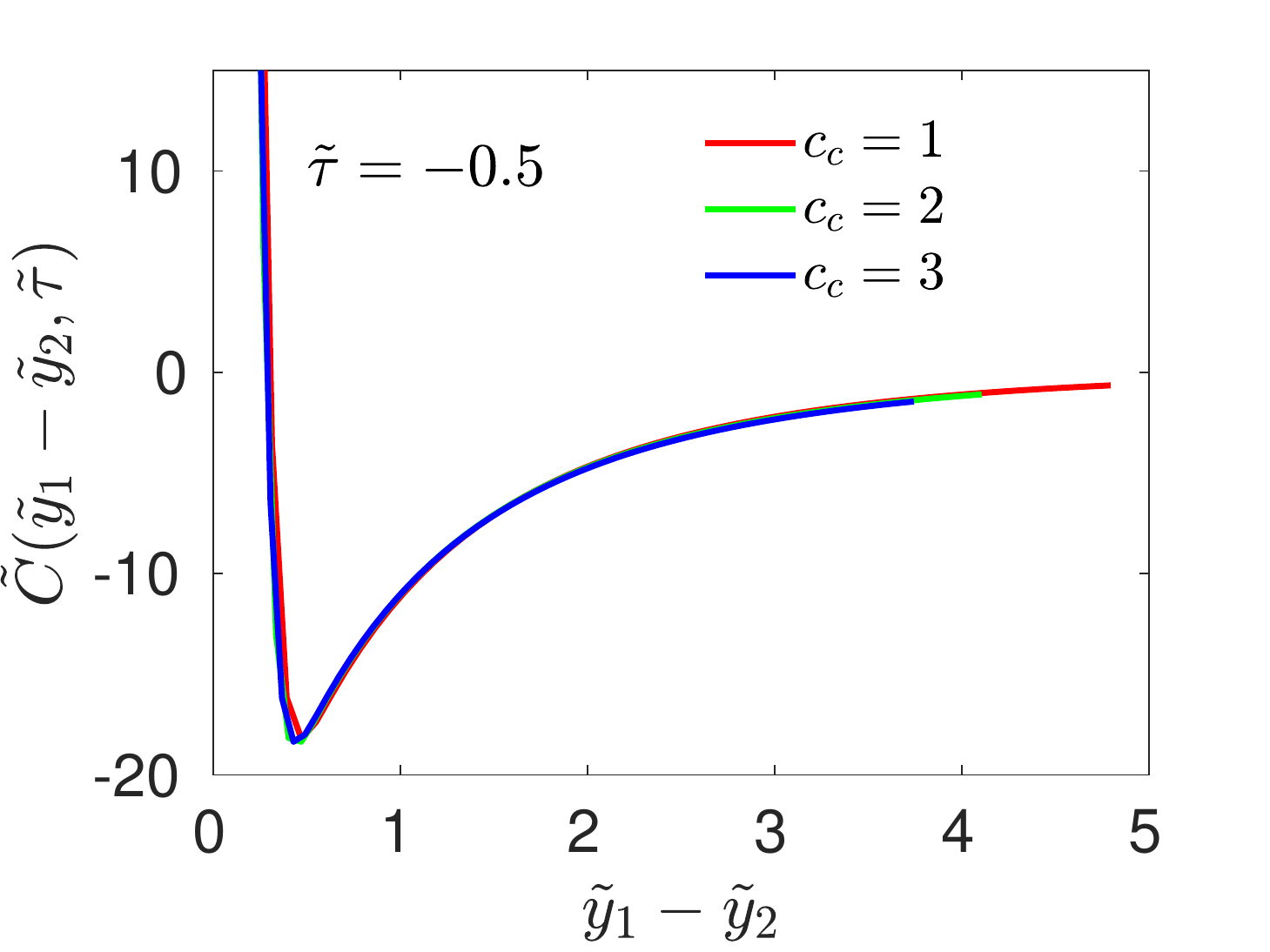}
\caption{(Color online) Left panel: the correlation function $C(y_1-y_2,\tau)$ of the conserved charge as a function of rapidity interval $y_1-y_2$, with different strength of critical component $c_c$. The rescaled time is fixed at $\tilde{\tau}=-0.5$ (where the temperature $T$ is close but above $T_c$). Right panel: the corresponding universal correlation function $\tilde{C}(\tilde{y}_1-\tilde{y}_2,\tilde{\tau})$ as a function of $\tilde{y}_1-\tilde{y}_2$.}
\label{Ccc-05}
\end{figure*}

\begin{figure*}[tbp]
\center
\includegraphics[width=3.3 in]{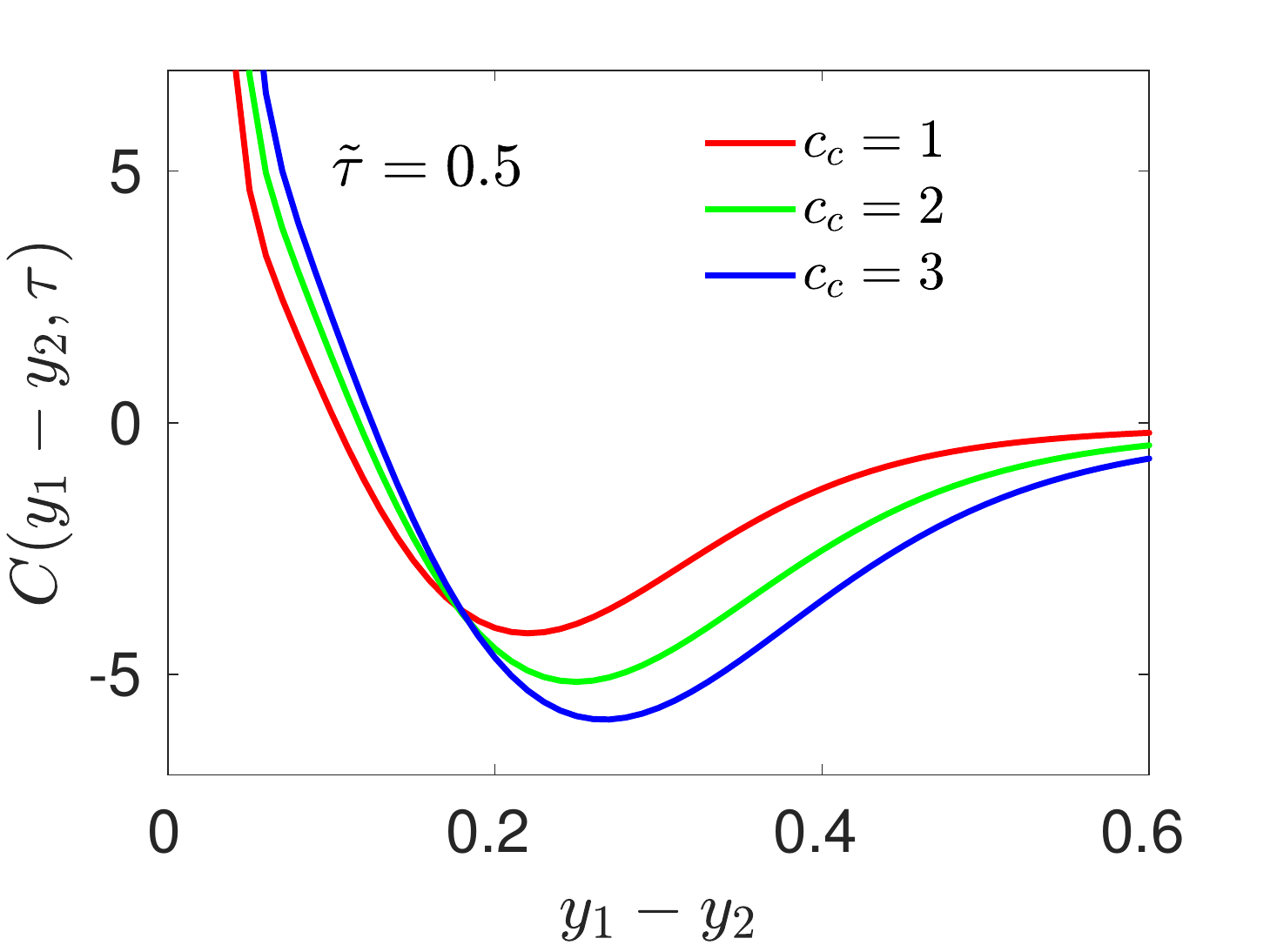}
\includegraphics[width=3.3 in]{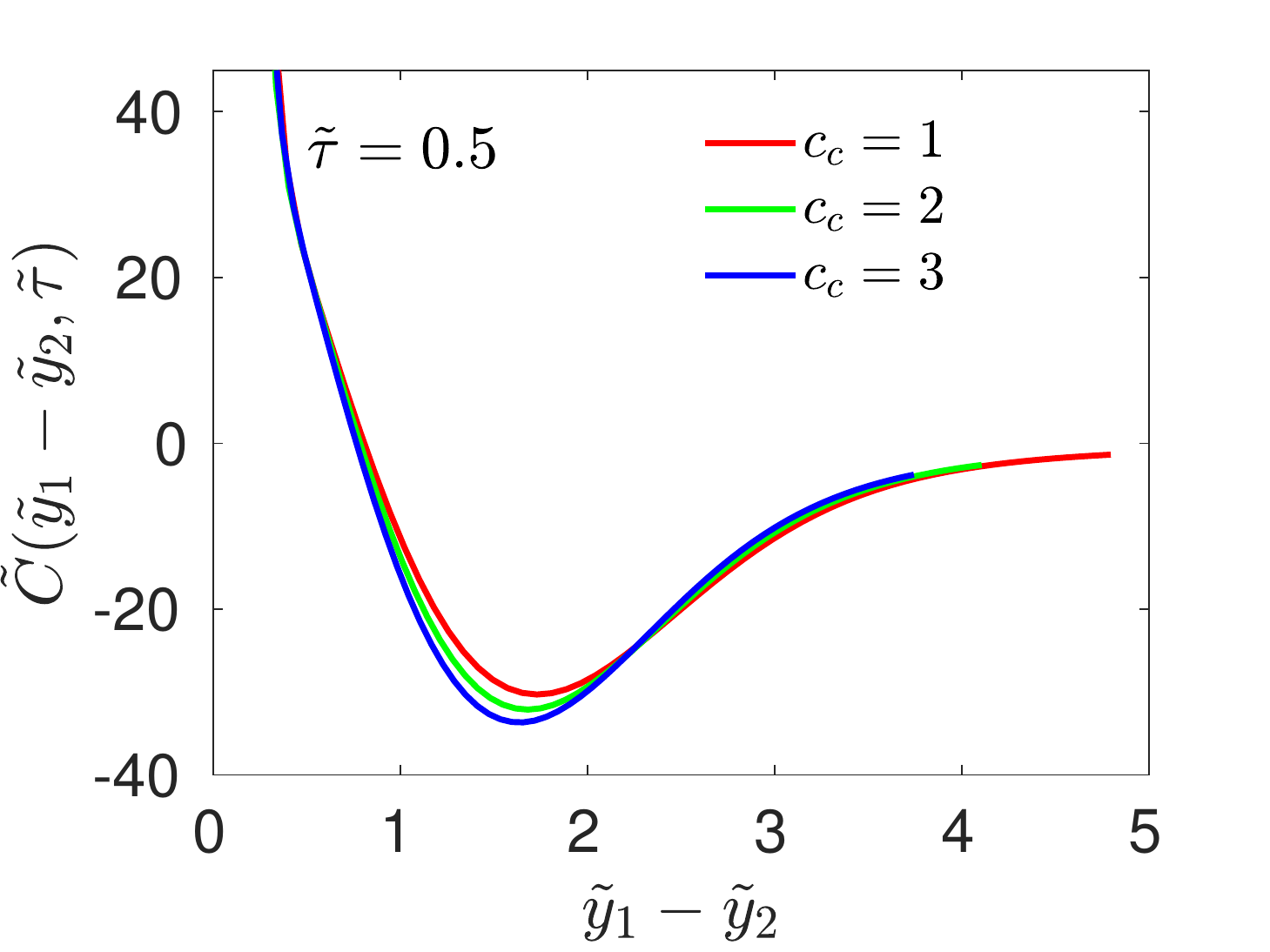}
\caption{(Color online) Similar to Fig.~\ref{Ccc-05}, {but with the rescaled time is set to $\tilde{\tau}=0.5$  (where the temperature $T$ is close but below $T_c$)}. }
\label{Ccc05}
\end{figure*}

After obtained the characteristic scales $\tau_{\mbox{\tiny KZ}},\,l_{\mbox{\tiny KZ}}$ in Eq.~(\ref{KZ}), one can construct the universal function with the following redefined variables:
\begin{align}\label{redefined}
\begin{aligned}
\tilde{\tau}&\equiv (\tau-\tau_c)/\tau_{\mbox{\tiny KZ}},\quad \tilde{y}\equiv y/l_{\mbox{\tiny KZ}},\quad \tilde{\xi}\equiv \xi/l_{\mbox{\tiny KZ}},&\\
\tilde{D}&\equiv D/l^{-2+\chi_\eta+\chi_\lambda}_{\mbox{\tiny KZ}},\quad \tilde{\chi} \equiv \chi /^{2-\chi_\eta}_{\mbox{\tiny KZ}}.&
\end{aligned}
\end{align}
For example, the rescaled correlation function $\tilde{C}(\tilde{y}_1-\tilde{y}_2,\tilde{\tau})$ and
the rescaled function of cumulant $\tilde{K}\left(\Delta y/l_{\mbox{\tiny KZ}}, \tilde{\tau}\right)$ can be
constructed as
\begin{align}\label{KZC}
\begin{aligned}
&C(y_1-y_2,\tau) \\
&=  l^{1-\chi_\eta}_{\mbox{\tiny KZ}}\Bigg\{ \tilde{\chi}(\tilde{\tau}) \delta (\tilde{y}_1-\tilde{y}_2)\\
  &\quad-\int^{\tilde{\tau}}_{\tilde{\tau}_0}d\tilde{\tau}'\frac{d\tilde{\chi}(\tilde{\tau}')}{d\tilde{\tau}'}[2\pi\int^{\tilde{\tau}}_{\tilde{\tau}'}d\tilde{\tau}''\tilde{D}]^{-\frac{1}{2}}\exp[-\frac{(\tilde{y}_1-\tilde{y}_2)^2}{2\int^{\tilde{\tau}}_{\tilde{\tau}'}d\tilde{\tau}''\tilde{D}}]\Bigg\}\\
  &\equiv  l^{1-\chi_\eta}_{\mbox{\tiny KZ}} \tilde{C}(\tilde{y}_1-\tilde{y}_2,\tilde{\tau}),
\end{aligned}
\end{align}
\begin{align}\label{KZK}
\begin{aligned}
K(\Delta y,\tau) &= l^{2-\chi_\eta}_{\mbox{\tiny KZ}} \Bigg\{\tilde{\chi}(\tilde{\tau})-\int^{\tilde{\tau}}_{\tilde{\tau}_0} d\tilde{\tau}'\frac{d\tilde{\chi}(\tilde{\tau}')}{d\tilde{\tau}'}F\bigg(\frac{\Delta y/l_{\mbox{\tiny KZ}}}{2[2\int^{\tilde{\tau}}_{\tilde{\tau}'}d\tilde{\tau}'' \tilde{D}]^{1/2}}\bigg)\Bigg\}\\
&\equiv l^{2-\chi_\eta}_{\mbox{\tiny KZ}} \tilde{K}\left(\frac{\Delta y}{l_{\mbox{\tiny KZ}}}, \tilde{\tau}\right).
\end{aligned}
\end{align}

The rescaled functions  $\tilde{C}(\tilde{y}_1-\tilde{y}_2,\tilde{\tau})$ and $\tilde{K}\left(\Delta y/l_{\mbox{\tiny KZ}}, \tilde{\tau}\right)$ as functions of the redefined variables $\tilde{y}_1-\tilde{y}_2,\tilde{\tau}$ and $\Delta y/l_{\mbox{\tiny KZ}}, \tilde{\tau}$ are universal and insensitive on the some free parameters, which will be demonstrated in the next section. {Note that the calculated correlation function $C(y_1-y_2,\tau)$  and cumulant $K(\Delta y,\tau)$ evolving with respect to proper time $\tau$, while the Kibble-Zurek scaling procedure is over the relative time $\tau-\tau_c$ as shown in Eq.~(\ref{KZ}). Therefore, the above rescaling formulae (\ref{KZC}) and (\ref{KZK}) valid near the critical point, where the relative time $\tau-\tau_c$ is small.}

\section{Results and discussions}\label{Results}
In this section, we will demonstrate the constructed universal functions Eq.~(\ref{KZC}) and Eq.~(\ref{KZK})  are insensitive to the free inputs, the strength of critical component $c_c$ and the initial temperature $T_0$.

Firstly, we numerically calculate the correlation function (\ref{Correl}) and cumulant (\ref{Cumul}) with the parameterizations of susceptibility $\chi(\tau)$ and diffusion coefficient $D(\tau)$ along a particular trajectory with the fixed chemical potential $r=0.1$. The temperature drops down according to Eq.~(\ref{Bjorken}) with the initial temperature $T_0=190$MeV and the initial time $\tau_0$ is set at: $\tilde{\tau}_0\equiv (\tau_0-\tau_c)/\tau_{\mbox{\tiny KZ}}=-2.5$.

{The left panel of Fig.~\ref{Ccc-05} presents the correlation function $C(y_1-y_2,\tau)$ as a function of $y_1-y_2$  with different strength of critical component $c_c=1,2,3$ at a fixed rescaled time $\tilde{\tau}=-0.5$, where the corresponding temperature $T$ is larger than but also close to $T_c$. As shown in Ref.~\cite{Sakaida:2017rtj}, the correlation function as a function of $y_1-y_2$ has a local minimum at very small $y_1-y_2$ due to the $\delta(y_1-y_2)$ contribution in Eq.~(\ref{Correl}).}
As expected, the correlation function~(\ref{Correl}) is sensitive to the strength of the critical component $c_c$.  In the right panel of Fig.~\ref{Ccc-05}, we investigate the universal behavior of the reconstructed correlation function~(\ref{KZC}) within the framework of KZM.  As shown in Fig.~\ref{KZcc},   the relaxation time $\tau_{\mbox{\tiny rel}}$ strongly enhances as the system approaches to the critical point and quench time $\tau_{\mbox{\tiny quench}}$ decreases, which results in a point $\tau^*$ where the relaxation time equals to the quench time.  With the obtained characteristic scales $l_{\mbox{\tiny KZ}}$ and $\tau_{\mbox{\tiny KZ}}$ at $\tau^*$ and the redefined variables~(\ref{redefined}), we construct the universal correlation function according to Eq.~(\ref{KZC}). The right panel of Fig.~\ref{Ccc-05} plots  the constructed universal correlation function $\tilde{C}(\tilde{y}_1-\tilde{y}_2,\tilde{\tau})$ at $\tilde{\tau}=-0.5$ with different $c_c$. Compared with the original  correlation function $C(y_1-y_2,\tau)$ that is sensitive to  critical component $c_c$, these constructed correlation function $\tilde{C}(\tilde{y}_1-\tilde{y}_2,\tilde{\tau})$ perfectly converge into one universal curve.

In Fig.~\ref{Ccc05} , we plot the correlation function $C(y_1-y_2,\tau)$ as a function of $y_1-y_2$ at the rescaled time $\tilde{\tau}=0.5$, where the temperature $T$ below $T_c$. { Different to the local minimum of $C(y_1-y_2,\tau)$ as a function of $y_1-y_2$ in Fig.~\ref{Ccc-05} which arises from the $\delta(y_1-y_2)$ contribution , the one in the left panel of Fig.~\ref{Ccc05} is due to the changing sign of $\chi'(\tau)$ in Eq.~(\ref{Correl}) when $T<T_c$ , indicating the susceptibility $\chi(\tau)$ has a maximum with respect to the proper time $\tau$.}  Meanwhile, $C(y_1-y_2,\tau)$ at $\tilde{\tau}=0.5$ also show sensitivity to the strength of the critical component $c_c$. After the same scaling procedure as above, the constructed universal correlation function $\tilde{C}(\tilde{y}_1-\tilde{y}_2,\tilde{\tau})$ nicely converge into one curve.

\begin{figure*}[tbp]
\center
\includegraphics[width=3.3 in]{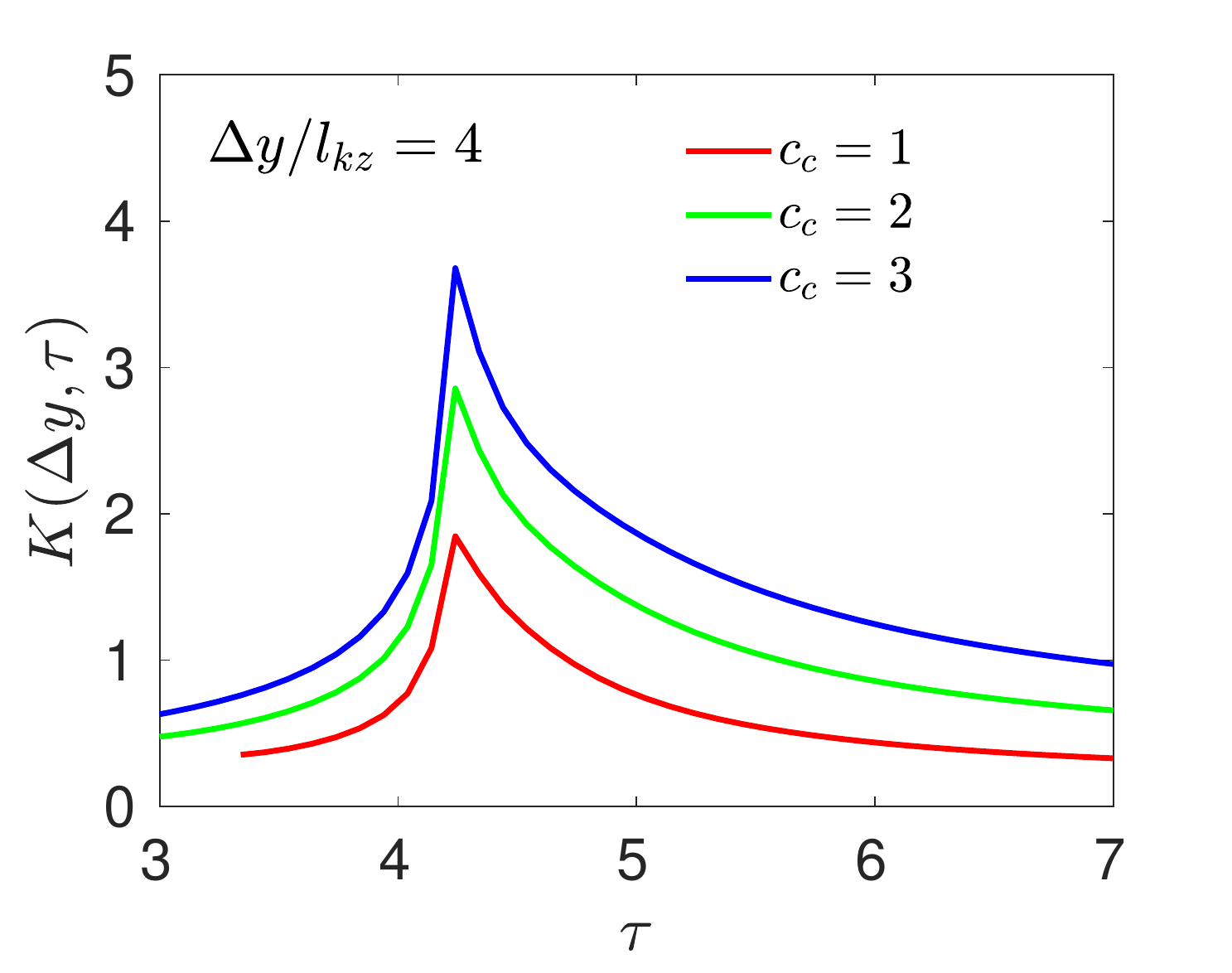}
\includegraphics[width=3.3 in]{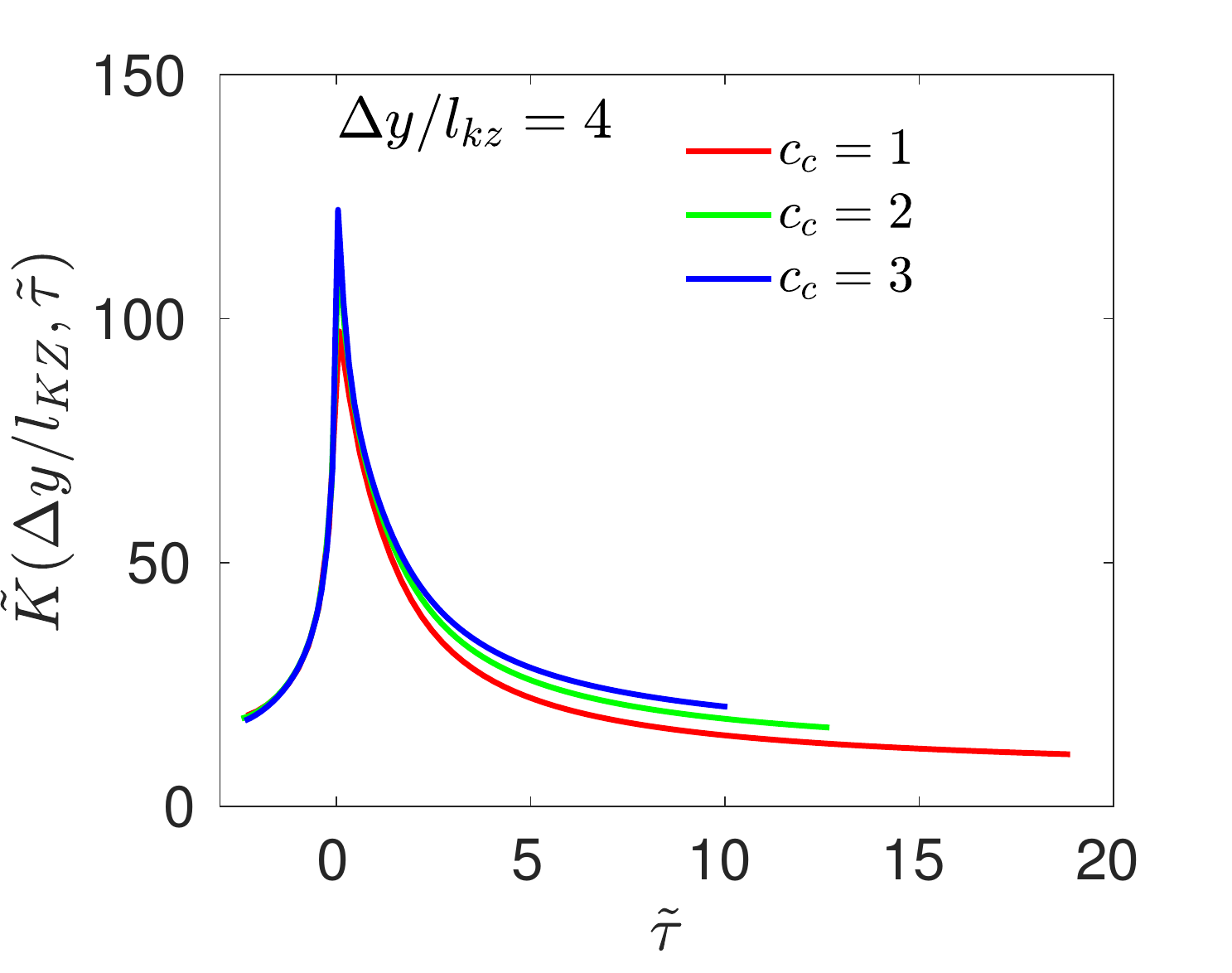}
\caption{(Color online) Time evolution of the second-order cumulants $K(\Delta y,\tau)$ for the conserved charge with different strength of critical component $c_c=1,2,3$. Right panel: the corresponding universal function $\tilde{K}(\Delta y/l_{\mbox{\tiny KZ}},\tilde{\tau})$ as a function of rescaled time $\tilde{\tau}$.}
\label{Kcc}
\end{figure*}
\begin{figure*}[tbp]
\center
\includegraphics[width=3.3 in]{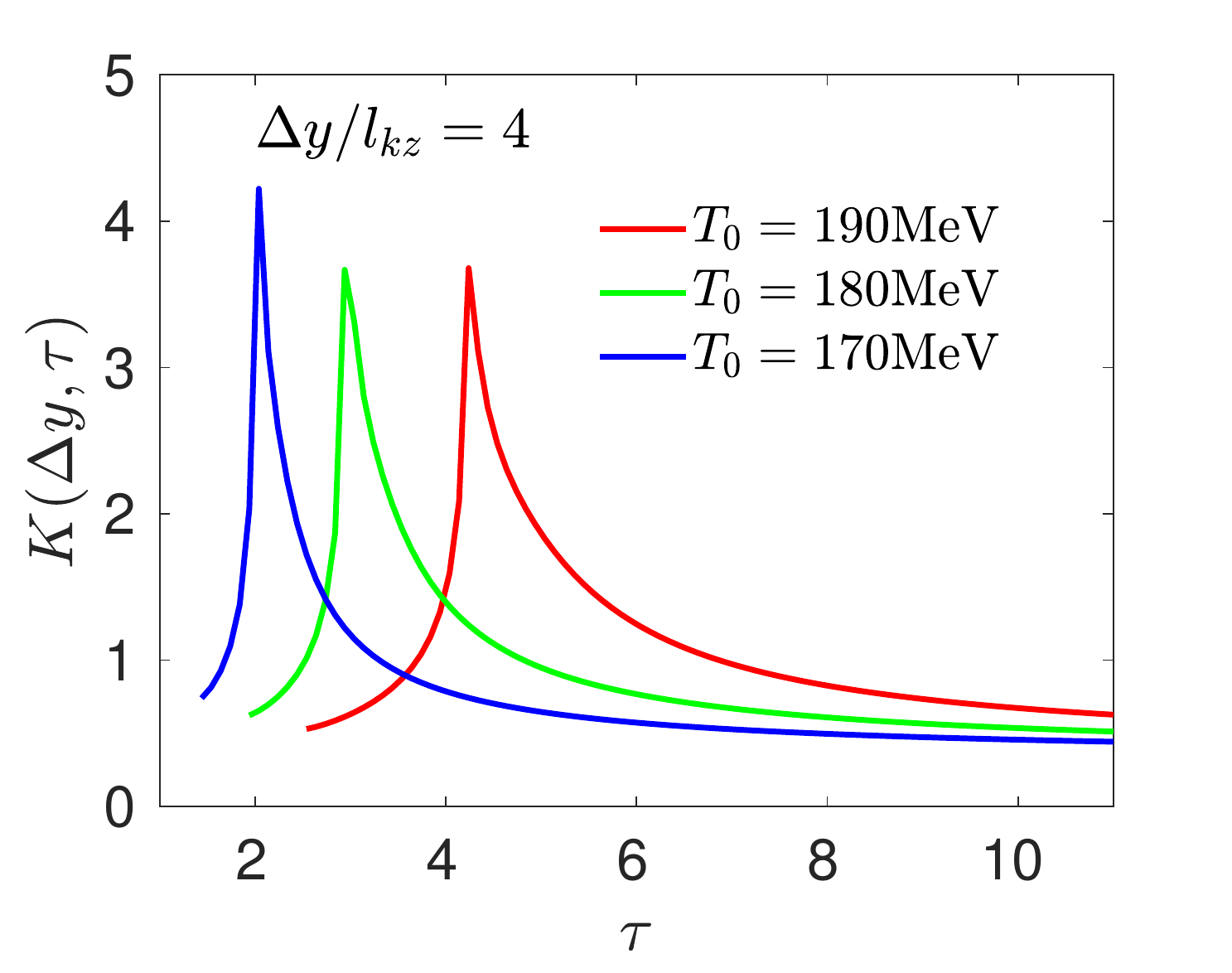}
\includegraphics[width=3.3 in]{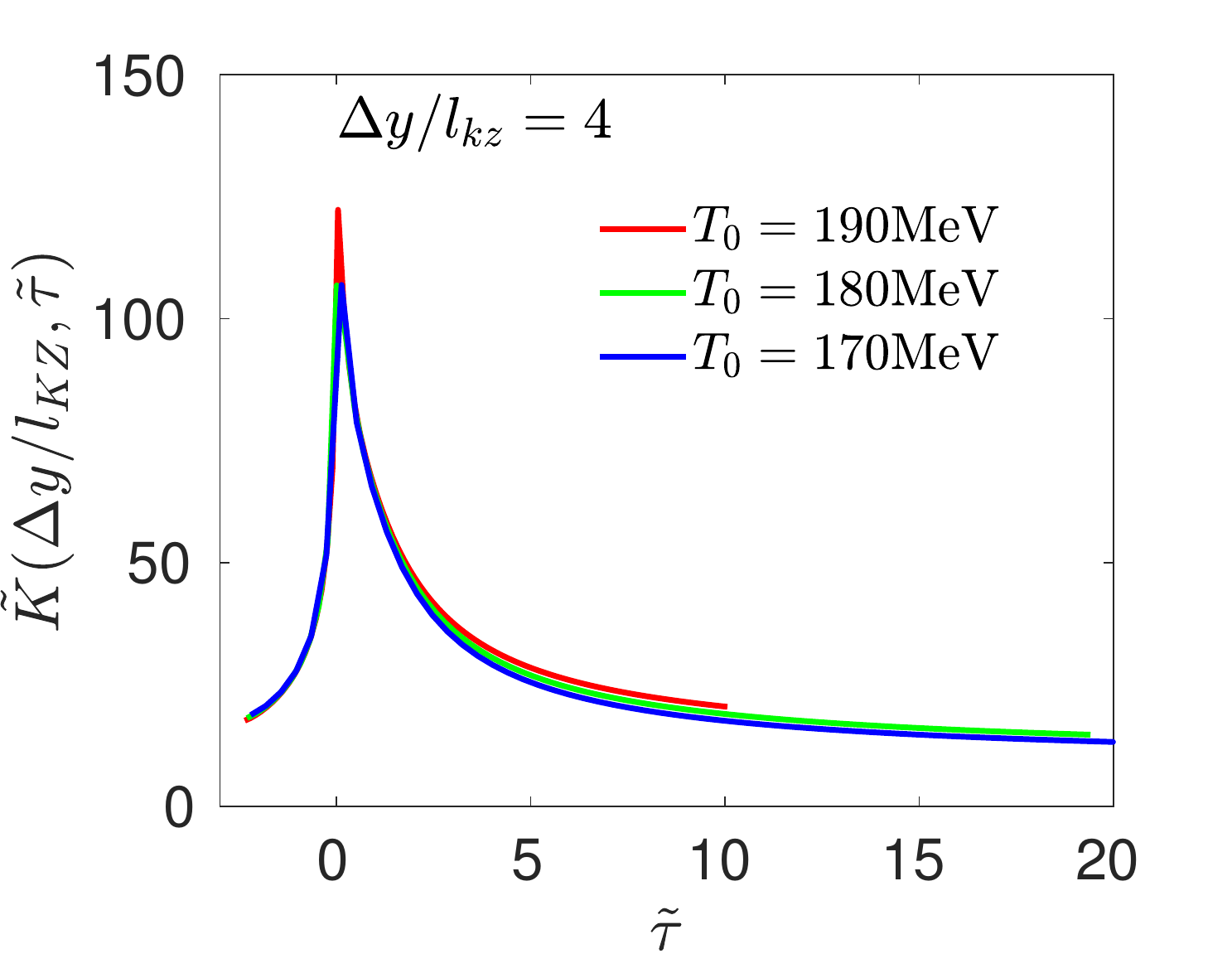}
\caption{(Color online) Similar to Fig.~4, but evolving the system with different initial temperatures $T_0=170,180,190$MeV.}
\label{KT0}
\end{figure*}

In Fig.~\ref{Kcc}, we investigate the universal behavior of the cumulant $K(\Delta y,\tau)$ according to Eq.~(\ref{KZK}). The system evolves with the same parameters as the two above cases, except for changing the chemical potential to $r=0.3$.  The left panel of Fig.~\ref{Kcc} shows the time evolution of $K(\Delta y,\tau)$ with different strengths of critical component $c_c$ , where $\Delta y$ is fixed at $\Delta y/l_{\mbox{\tiny KZ}}=4$.  Similar to the two above cases of correlation function, the time evolution of second-order cumulant  $K(\Delta y,\tau)$ strongly depends on $c_c$. After rescaling $K(\Delta y,\tau)$  and $\tau-\tau_c$  with $l^{2-\chi_\eta}_{\mbox{\tiny KZ}}$ and $\tau_{\mbox{\tiny KZ}}$, the constructed unviversal cumulant $\tilde{K}(\Delta y/l_{\mbox{\tiny KZ}},\tilde{\tau})$  is independent on the strength of the critical component $c_c$, as expected in Eq.~(\ref{KZK}).


In the last paragraph of this section, we will show that the constructed universal function $\tilde{K}(\Delta y/l_{\mbox{\tiny KZ}},\tilde{\tau})$  is also not sensitive to the initial temperature $T_0$. For this case, we evolve the systems with different initial temperature $T_0=170,\ 180,\ 190$ MeV at a {fixed initial rescaled} time $\tilde{\tau}_0=-2.5$ along a trajectory with fixed chemical potential $r=0.3$. Again, we assume one dimensional Bjorken expansion and the temperature drops according
to Eq.~(\ref{Bjorken}). The left panel of Fig.~\ref{KT0} plots the time evolution of the second-order cumulant $K(\Delta y,\tau)$ with $c_c=3$ and  $\Delta y/l_{\mbox{\tiny KZ}}=4$, which shows a significant dependence on the initial temperature $T_0$.  After the same rescaling procedure as described above, the universal cumulant $\tilde{K}(\Delta y/l_{\mbox{\tiny KZ}},\tilde{\tau})$ is constructed, which is insensitive to the initial temperature $T_0$ as shown  in right panel of Fig.~\ref{KT0}

\section{summary and outlook}\label{summary}

In this paper, we explored the Kibble-Zurek scaling for the critical fluctuation of the conserved charge within the framework of stochastic diffusion dynamics.  By analytically solving the stochastic diffusion equation~(\ref{SDE}), the time evolution of the two-point correlation function $C(y_1-y_2,\tau)$ and the second-order cumulant $K(\Delta y,\tau)$ of conserved charge are obtained, which are non-universal in terms of some free inputs in the model calculations, such as the initial temperature $T_0$ and the strengths of the critical components $c_c$ in the mapping between the QCD medium and 3D-Ising model.

With determinating the time $\tau^*$ after which the system falls out-of-equilibrium, we calculated the characteristic scales $\tau_{\mbox{\tiny KZ}}$ and $l_{\mbox{\tiny KZ}}$ of the ``frozen" system near the critical point. Using these obtained scales and rescaling the traditional two-point correlation function $C(y_1-y_2,\tau)$ and cumulant $K(\Delta y,\tau)$ , {we constructed the universal correlation function $\tilde{C}(\tilde{y}_1-\tilde{y}_2,\tilde{\tau})$ and cumulant $\tilde{K}(\Delta y/l_{\mbox{\tiny KZ}},\tilde{\tau})$ in terms of the rescaled rapidity $\tilde{y}$ and proper time $\tilde{\tau}$, respectively}.  These rescaled functions are universal in terms of different free parameters.  For instance, we have numerically   shown  that the universal functions  $\tilde{C}(\tilde{y}_1-\tilde{y}_2,\tilde{\tau})$ and $\tilde{K}(\Delta y/l_{\mbox{\tiny KZ}},\tilde{\tau})$ nicely converge into one curve which are insensitive to the strength of critical component $c_c$ and initial temperature $T_0$, respectively.

At last we would like to point out that this work focuses on the universal scaling of the two-point correlation function and second-order cumulant for the conserved charge based on the stochastic diffusion equation without the higher order coupling~(\ref{SDE}). At current stage, one can not also expect to connect  our constructed universal functions with the experimental data since we used the 1+1-dimensional heat bath with Bjorken approximation to simplify the calculations. On the other hand, there are many natural extensions to this current study. For example, with the higher order contribution added to the stochastic diffusion equation of the conserved charge, one can not only study the universal scaling of the two-point correlation function, but also the ones of  multi-point correlation functions and related higher-order cumulants.  Besides, studying the universal scaling with a more realistic evolving medium are also important for a realistic predictions of the possible observable that might be measured in experiment.  These work are complicated, but worthwhile to be investigated in the near future.

\section*{Acknowledgements}
We would like to thank the fruitful discussion with F.~Yan,  D.~Teaney, M.~Kitazawa and M.~Asakawa. This work is supported by the NSFC and the MOST under grant Nos. 11675004, 11435001 and 2015CB856900. We also gratefully acknowledge the extensive computing resources provided by the Super-computing Center of Chinese Academy of Science (SCCAS), Tianhe-1A from the National Supercomputing Center in Tianjin, China and the High-performance Computing Platform of Peking University.

\appendix

\section{Derivation for the time evolution of correlation function}\label{Appendix}

In this appendix, we presents the detail derivation of correlation function~(\ref{Correl}) from the stochastic diffusion equation~(\ref{SDE}).

With the Fourier transform
\begin{align}
n(q,\tau) = \int dy e^{-iqy}n(y,\tau),
\end{align}
SDE~(\ref{SDE}) in the Fourier space is written as: 
\begin{align}
\frac{\partial}{\partial \tau} \delta n(q,\tau) = -D(\tau) q^2 \delta n(q,\tau) + iq \zeta(q,\tau),
\end{align}
and the noise satisfies
\begin{align}
\begin{aligned}
&\langle \zeta(q,\tau) \rangle =0,\\
&\langle \zeta(q_1,\tau_1) \zeta(q_2,\tau_2) \rangle = 4\pi \chi(\tau) D(\tau) \delta (q_1+q_2) \delta (\tau_1-\tau_2).
\end{aligned}
\end{align}
Therefore, one could obtain  the time evolution of correlation function in $q$ space :
\begin{align}\label{eqCorr}
\begin{aligned}
\frac{\partial}{\partial \tau}&\langle \delta n(q_1,\tau) \delta n(q_2,\tau) \rangle\\
& = -D(\tau) (q^2_1+q^2_2) \langle \delta n(q_1,\tau) \delta n(q_2,\tau) \rangle\nonumber
\\&\quad + 4\pi q_1q_2 \chi(\tau) D(\tau) \delta (q_1+q_2),
\end{aligned}
\end{align}
based on which the relaxation time of the correlation function is obtained as: $ \tau_{\mbox{\tiny rel}} = [ D(\tau) (q^2_1+q^2_2)]^{-1}$. With the assumption of the locality in the initial fluatuation
\begin{align}
\langle \delta n(q_1,\tau_0) \delta n(q_2,\tau_0) \rangle=2\pi \delta(q_1+q_2) \chi(\tau_0),
\end{align}
the solution of Eq.~(\ref{eqCorr}) is calculated to be
\begin{align}
\begin{aligned}
&\langle \delta n(q_1,\tau) \delta n(q_2,\tau) \rangle\\
&=2\pi \delta(q_1+q_2) \bigg (\chi(\tau_0) e^{-q^2_1[d(\tau_0,\tau)]^2}\\
&\quad+2q^2_1\int^\tau_{\tau_0} d\tau' \chi(\tau')D(\tau')e^{-q^2_1[d(\tau',\tau)]^2}\bigg).
\end{aligned}
\end{align}
Then, the correlation function in $y$ space is computed as
\begin{align}
\begin{aligned}
&\langle \delta n(y_1,\tau) \delta n(y_2,\tau) \rangle\\
&=\chi(\tau_0) G(y_1-y_2;2d(\tau_0,\tau))\\
&\qquad +\int^\tau_{\tau_0} d\tau'\chi(\tau') \frac{d}{d\tau'} G(y_1-y_2;2d(\tau',\tau))\\
&= \chi(\tau) \delta(y_1-y_2)\\
 &\qquad- \int^\tau_{\tau_0} d\tau' \chi'(\tau') G(y_1-y_2;2d(\tau',\tau)).
\end{aligned}
\end{align}
Meanwhile, the second order cumulant $K(\Delta y,\tau)$ can straightforwardly calculated, as shown in Eq.~(\ref{Cumul}).


\begin{thebibliography}{99}
\bibitem{Aggarwal:2010cw}
  M.~M.~Aggarwal {\it et al.} [STAR Collaboration],
  arXiv:1007.2613 [nucl-ex].

\bibitem{Mohanty:2011nm}
  B.~Mohanty [STAR Collaboration],
  J.\ Phys.\ G {\bf 38}, 124023 (2011).

\bibitem{Kumar:2013cqa}
  L.~Kumar,
  Mod.\ Phys.\ Lett.\ A {\bf 28}, 1330033 (2013).

\bibitem{Adamczyk:2017iwn}
  L.~Adamczyk {\it et al.} [STAR Collaboration],
  Phys.\ Rev.\ C {\bf 96}, 044904 (2017).

\bibitem{Odyniec:2015iaa}
  G.~Odyniec,
  EPJ Web Conf.\  {\bf 95}, 03027 (2015).

\bibitem{Stephanov:1999zu}
  M.~A.~Stephanov, K.~Rajagopal and E.~V.~Shuryak,
  Phys.\ Rev.\ D {\bf 60}, 114028 (1999).

\bibitem{Stephanov:1998dy}
  M.~A.~Stephanov, K.~Rajagopal and E.~V.~Shuryak,
  Phys.\ Rev.\ Lett.\  {\bf 81}, 4816 (1998).

\bibitem{Stephanov:2004wx}
  M.~A.~Stephanov,
  Prog.\ Theor.\ Phys.\ Suppl.\  {\bf 153}, 139 (2004).

\bibitem{Stephanov:2007fk}
  M.~A.~Stephanov,
  PoS LAT {\bf 2006}, 024 (2006).

\bibitem{Asakawa:2015ybt}
  M.~Asakawa and M.~Kitazawa,
  Prog.\ Part.\ Nucl.\ Phys.\  {\bf 90}, 299 (2016).

\bibitem{Luo:2017faz}
  X.~Luo and N.~Xu,
  Nucl.\ Sci.\ Tech.\  {\bf 28}, no. 8, 112 (2017).

\bibitem{Li:2017ple}
  Z.~Li, Y.~Chen, D.~Li and M.~Huang,
  Chin.\ Phys.\ C {\bf 42}, no. 1, 013103 (2018)
  
\bibitem{Chen:2018vty} 
  X.~Chen, D.~Li and M.~Huang,
  arXiv:1810.02136 [hep-ph].
  
\bibitem{Li:2018ygx} 
  Z.~Li, K.~Xu, X.~Wang and M.~Huang,
  arXiv:1801.09215 [hep-ph].
  
\bibitem{Fu:2015amv} 
  W.~j.~Fu and J.~M.~Pawlowski,
  Phys.\ Rev.\ D {\bf 93}, 091501 (2016)


\bibitem{Klevansky:1992qe}
  S.~P.~Klevansky,
  Rev.\ Mod.\ Phys.\  {\bf 64}, 649 (1992).

\bibitem{Fukushima:2003fw}
  K.~Fukushima,
  Phys.\ Lett.\ B {\bf 591}, 277 (2004).

\bibitem{Fu:2007xc}
  W.~j.~Fu, Z.~Zhang and Y.~x.~Liu,
  Phys.\ Rev.\ D {\bf 77}, 014006 (2008).

\bibitem{Jiang:2013yoa}
  L.~j.~Jiang, X.~y.~Xin, K.~l.~Wang, S.~x.~Qin and Y.~x.~Liu,
  Phys.\ Rev.\ D {\bf 88}, 016008 (2013).

\bibitem{Roberts:1994dr}
  C.~D.~Roberts and A.~G.~Williams,
  Prog.\ Part.\ Nucl.\ Phys.\  {\bf 33}, 477 (1994).

\bibitem{Qin:2010nq}
  S.~x.~Qin, L.~Chang, H.~Chen, Y.~x.~Liu and C.~D.~Roberts,
  Phys.\ Rev.\ Lett.\  {\bf 106}, 172301 (2011).

\bibitem{Berges:2000ew}
  J.~Berges, N.~Tetradis and C.~Wetterich,
  Phys.\ Rept.\  {\bf 363}, 223 (2002).


\bibitem{Stephanov:2008qz} M.~A.~Stephanov,
Phys.\ Rev.\ Lett.\ \textbf{102}, 032301 (2009).





\bibitem{Aggarwal:2010wy} M.~M.~Aggarwal \textit{et al.} [STAR
Collaboration],
Phys.\ Rev.\ Lett.\ \textbf{105}, 022302 (2010).

\bibitem{Adamczyk:2013dal} L.~Adamczyk \textit{et al.} [STAR Collaboration],
Phys.\ Rev.\ Lett.\ \textbf{112}, 032302 (2014).


\bibitem{Adamczyk:2014fia}
  L.~Adamczyk {\it et al.} [STAR Collaboration],
  Phys.\ Rev.\ Lett.\  {\bf 113}, 092301 (2014).
\bibitem{Thader:2016gpa}
  J.~Th$\ddot{\mbox{a}}$der [STAR Collaboration],
  Nucl.\ Phys.\ A {\bf 956}, 320 (2016).

 \bibitem{Luo:2015ewa} X.~Luo [STAR Collaboration],
PoS CPOD \textbf{2014}, 019 (2014). 

\bibitem{Stephanov:2011pb} M.~A.~Stephanov,
Phys.\ Rev.\ Lett.\ \textbf{107}, 052301 (2011).





\bibitem{Rajagopal:2000}
  B.~Berdnikov and K.~Rajagopal,
  Phys.\ Rev.\ D {\bf 61}, 105017 (2000)

\bibitem{Paech:2003fe} K.~Paech, H.~Stoecker and A.~Dumitru,
Phys.\ Rev.\ C \textbf{68}, 044907 (2003). 


\bibitem{Son:2004iv} D.~T.~Son and M.~A.~Stephanov,
Phys.\ Rev.\ D \textbf{70}, 056001 (2004).

\bibitem{Nonaka:2004pg}
  C.~Nonaka and M.~Asakawa,
  Phys.\ Rev.\ C {\bf 71}, 044904 (2005).

\bibitem{Asakawa:2009aj}
  M.~Asakawa, S.~Ejiri and M.~Kitazawa,
  Phys.\ Rev.\ Lett.\  {\bf 103}, 262301 (2009).

\bibitem{Stephanov:2009ra}
  M.~A.~Stephanov,
  Phys.\ Rev.\ D {\bf 81}, 054012 (2010).


\bibitem{Nahrgang:2011mg} M.~Nahrgang, S.~Leupold, C.~Herold and
M.~Bleicher,
Phys.\ Rev.\ C \textbf{84}, 024912 (2011). 


  \bibitem{Mukherjee:2015swa} S.~Mukherjee, R.~Venugopalan and Y.~Yin,
Phys.\ Rev.\ C \textbf{92}, 034912 (2015).

\bibitem{Jiang:2017mji}
  L.~Jiang, S.~Wu and H.~Song,
  Nucl.\ Phys.\ A {\bf 967}, 441 (2017).

\bibitem{Ling:2015yau}
  B.~Ling and M.~A.~Stephanov,
  Phys.\ Rev.\ C {\bf 93}, 034915 (2016).


\bibitem{Jiang:2015hri} L.~Jiang, P.~Li and H.~Song,
Phys.\ Rev.\ C \textbf{94}, 024918 (2016);
Nucl.\ Phys.\ A {\bf 956}, 360 (2016).


\bibitem{Mukherjee:2016kyu} S.~Mukherjee, R.~Venugopalan and Y.~Yin,
Phys.\ Rev.\ Lett.\ \textbf{117}, 222301 (2016).






\bibitem{Sakaida:2017rtj}
  M.~Sakaida, M.~Asakawa, H.~Fujii and M.~Kitazawa,
  Phys.\ Rev.\ C {\bf 95},  064905 (2017).

\bibitem{Brewer:2018abr}
  J.~Brewer, S.~Mukherjee, K.~Rajagopal and Y.~Yin,
  arXiv:1804.10215 [hep-ph].

\bibitem{Stephanov:2017ghc}
  M.~Stephanov and Y.~Yin,
  Phys.\ Rev.\ D {\bf 98},  036006 (2018).





\bibitem{Nahrgang:2018afz}
  M.~Nahrgang, M.~Bluhm, T.~Schaefer and S.~A.~Bass,
  arXiv:1804.05728 [nucl-th].

\bibitem{Akamatsu:2018vjr} 
  Y.~Akamatsu, D.~Teaney, F.~Yan and Y.~Yin,
  arXiv:1811.05081 [nucl-th].









\bibitem{Chandran:2012}
A.~Chandran, A.~Erez, S.~S.~Gubser, S.~L.~Sondhi,
  Phys.\ Rev.\ B {\bf 86}, 064304 (2012).

\bibitem{Kolodrubetz:2012}
M.~Kolodrubetz, B.~K.~Clark, D.~A.~Huse,
  Phys.\ Rev.\ Lett.\  {\bf 109}, 015701 (2012).

\bibitem{Francuz:2015zva}
  A.~Francuz, J.~Dziarmaga, B.~Gardas and W.~H.~Zurek,
  Phys.\ Rev.\ B {\bf 93}, 075134 (2016).

\bibitem{Nikoghosyan:2013fqa}
  G.~Nikoghosyan, R.~Nigmatullin and M.~B.~Plenio,
  Phys.\ Rev.\ Lett.\  {\bf 116}, 080601 (2016).

\bibitem{Wu:2018twy}
  S.~Wu, Z.~Wu and H.~Song,
  arXiv:1811.09466 [nucl-th].
  



\bibitem{Kibble:1976}T.~W.~B.~Kibble,
J.\ Phys.\ A:\ Math. \ Gen. \textbf{9}, 1387 (1976).

\bibitem{Zurek:1985}W.~H.Zurek,
Nature \textbf{317}, 505 (1985).



\bibitem{Rev1977}
P.~C.~Hohenberg and B.~I.~Halperin,
Rev.\ Mod. \ Phys. \textbf{49}, 435 (1977).

%

\bibitem{Fujii:2003bz}
  H.~Fujii,
  Phys.\ Rev.\ D {\bf 67}, 094018 (2003).
\bibitem{Fujii:2004za}
  H.~Fujii and M.~Ohtani,
  Prog.\ Theor.\ Phys.\ Suppl.\  {\bf 153}, 157 (2004).
\bibitem{Fujii:2004jt}
  H.~Fujii and M.~Ohtani,
  Phys.\ Rev.\ D {\bf 70}, 014016 (2004).




\bibitem{Justin:2001}J.~Zinn-Justin,
Phys.\ Rept.\ \textbf{344},159 (2001).

\bibitem{Schofield:1969}P.~Schofield, J.~D.~Litster, and J.~T.~Ho,
Phys.\ Rev.\ Lett.\ \textbf{23},1098 (1969).

\bibitem{Guida:1996ep}
  R.~Guida and J.~Zinn-Justin,
  Nucl.\ Phys.\ B {\bf 489}, 626 (1997).



\end{thebibliography}
\end{document}